\documentclass[fdp,a4paper,fleqn,finallayout]{w-art}
\usepackage{times,cite,w-thm}
\usepackage[]{graphicx}

\begin{document}

\DOIsuffix{theDOIsuffix}
\Volume{55}
\Issue{1}
\Month{01}
\Year{2007}
\pagespan{1}{}
\Receiveddate{}
\Reviseddate{}
\Accepteddate{}
\Dateposted{}

\title[Observational Effects of Strong Gravity in Vicinity of Supermassive Black Holes]
{Observational Effects of Strong Gravity in Vicinity of \\
Supermassive Black Holes}

\author[P. Jovanovi\' c]{Predrag Jovanovi\' c\inst{1,}%
  \footnote{Corresponding author\quad E-mail:~\textsf{pjovanovic@aob.bg.ac.yu},
            Phone: +381\,11\,3089\,068,
            Fax: +381\,11\,2419\,553}}
\address[\inst{1}]{Astronomical Observatory, Volgina 7, 11160 Belgrade, Serbia}

\author[L. \v C. Popovi\' c]{Luka \v C. Popovi\' c\inst{1,2}}
\address[\inst{2}]{Alexander von Humboldt Fellow, presently at Max Planck Institute for Radioastronomy, Bonn, Germany}

\begin{abstract}
Here we discuss the effects of strong gravity that can be observed
in electromagnetic spectra of active galactic nuclei (AGN).
According to the unification model of an AGN, there is a
supermassive black hole ($10^7 - 10^9 M_\odot$) in its center,
surrounded by an accretion disk that radiates in the X-ray band.
Accretion disks could have different forms, dimensions, and
emission, depending on the type of central black hole (BH), whether
it is rotating (Kerr metric) or nonrotating (Schwarzschild metric).
We modeled the emission of an accretion disk around supermassive BH
using numerical simulations based on a ray-tracing method in the
Kerr metric. A broad emission line Fe K$\alpha$ at 6.4 keV with
asymmetric profile (narrow bright blue peak and a wide faint red
wing) has been observed in a number of type 1 AGN. The effects of
strong gravitational field are investigated by comparison between
the modeled and observed iron K$\alpha$ line profiles. The results
of our modeling show that the parameters of the Fe K$\alpha$ line
emitting region have significant influence on the line profile and
thus, allow us to determine the space-time geometry (metric) in
vicinity of the central BH of AGN, and also can give us information
about the plasma conditions in these regions.
\end{abstract}

\maketitle

\section{Introduction}

It is now widely accepted that AGN derive their extraordinary
luminosities (sometimes more than $10^4$ times higher than
luminosities of "ordinary" galaxies) from energy release by matter
accreting towards, and falling into, a central supermassive BH. The
accretion disks around the central BH represent an efficient
mechanism for extracting gravitational potential energy and
converting it into radiation, giving us the most probable
explanation for the main characteristics of AGN (high luminosity,
compactness, jet formation, rapid time variation in radiation and
the profile of the Fe K$\alpha$ spectral line). Thus, AGN are
powerful sources of radiation in a wide spectral range: from
$\gamma$ rays to radio waves \cite{kr99}.

The most important feature of the X-ray radiation of AGN (which is
generated in the innermost region around a central BH) is a broad
emission line Fe K$\alpha$ at 6.4 keV that may have an asymmetric
profile (narrow bright blue peak and wide faint red peak). It was
discovered in Seyfert 1 galaxy MCG-6-30-15 \cite{tan95} and later on
observed in a number of AGN. In some cases the line width
corresponds to one third of speed of light, indicating that its
emitters rotate with relativistic velocities. Therefore, the line is
probably produced in a very compact region near the central BH of
AGN and can provide us some essential information about the plasma
conditions and the space-time geometry in vicinity of the BH
\cite{pop03a}.

Black holes have only three measurable parameters (not including the
Hawking temperature): charge, mass and angular momentum
\cite{yaq07}. In this paper we will pay attention mostly to angular
momentum or spin of central supermassive BH of AGN, which is a
property of the space-time metric.

\begin{figure}[b!]
\centering
\includegraphics[width=0.8\textwidth]{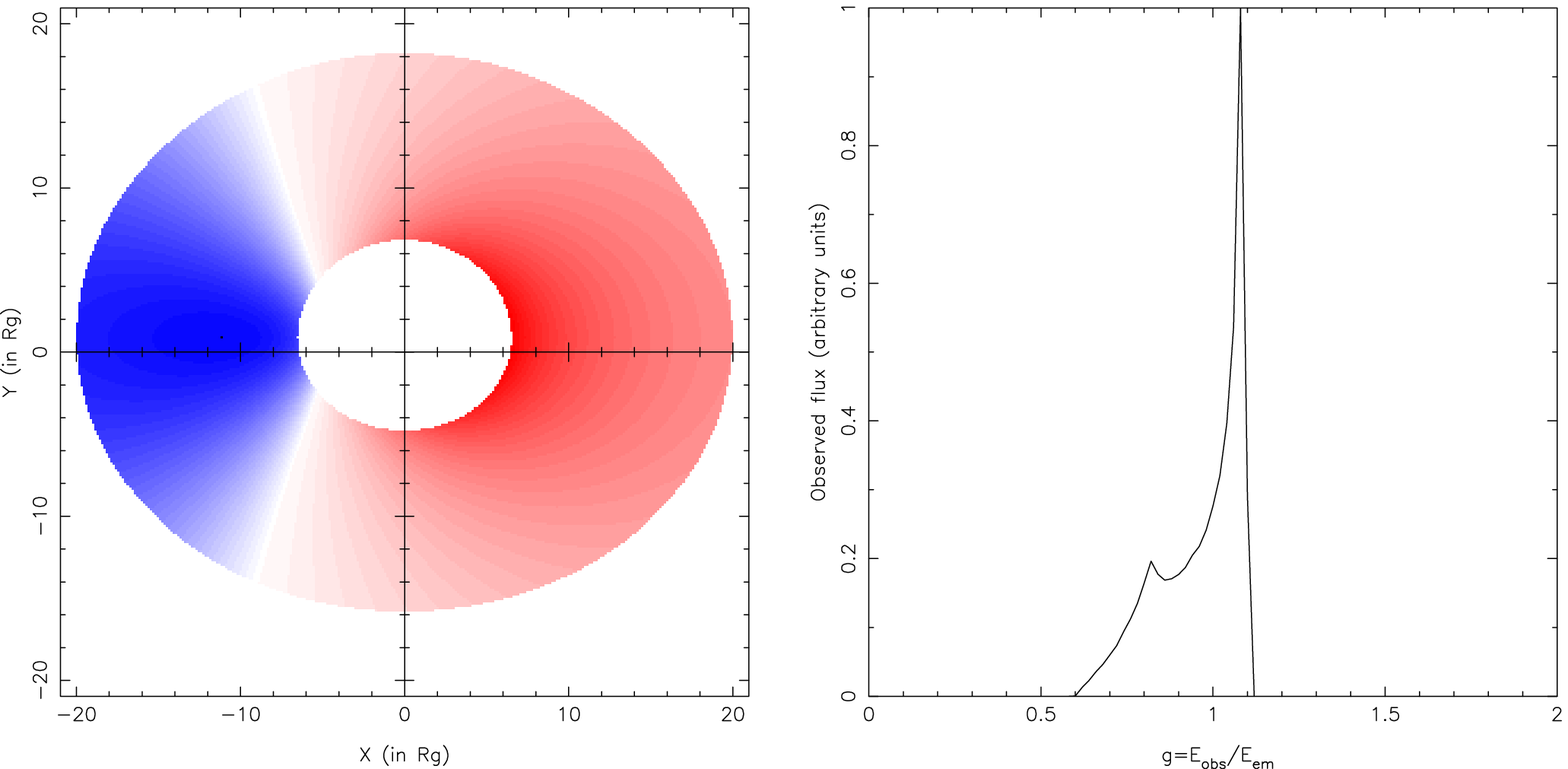} \\
\includegraphics[width=0.8\textwidth]{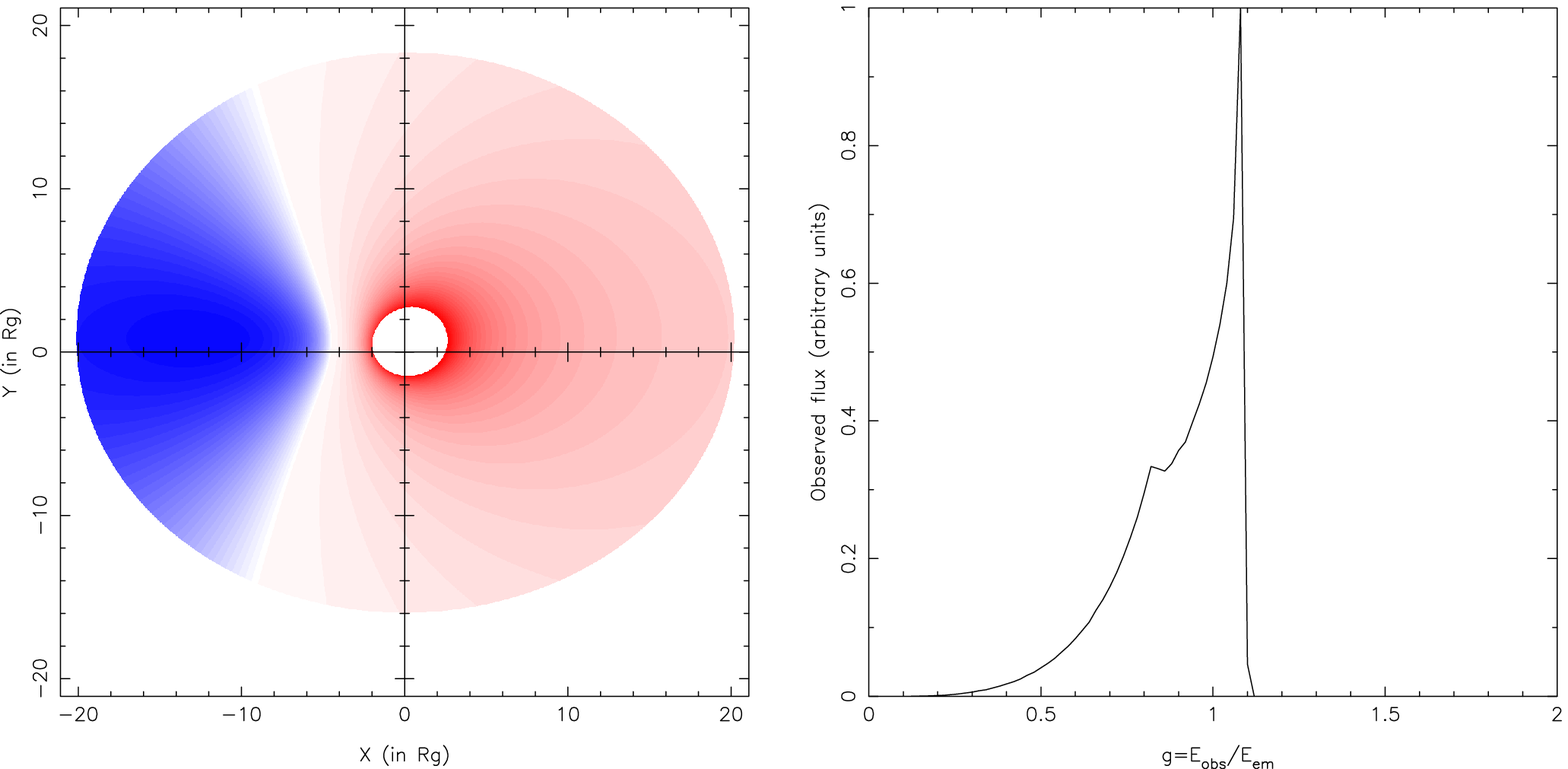}
\caption{Illustrations of accretion disk (left) and the
corresponding Fe K$\alpha$ line profiles (right) in the case of
Schwarzschild (top) and Kerr metric with angular momentum parameter
$a=0.998$ (bottom). The disk inclination is $i=35^\circ$ and its
inner and outer radii are $R_{in}=R_{ms}$ and $R_{out}=20$ R$_{g}$,
respectively.} \label{fig1}
\end{figure}

\begin{figure}[ht!]
\centering
\includegraphics[width=0.8\textwidth]{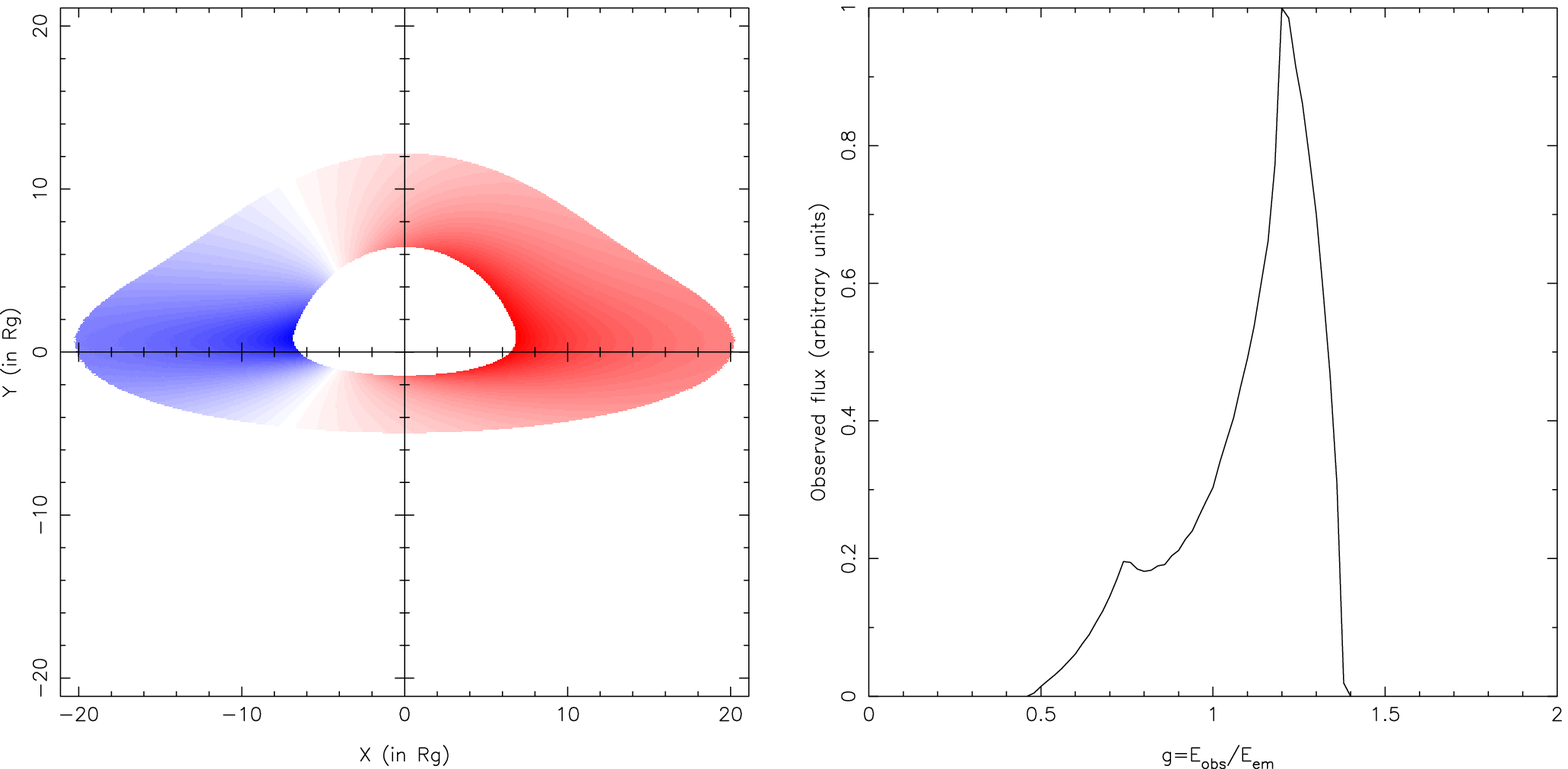} \\
\includegraphics[width=0.8\textwidth]{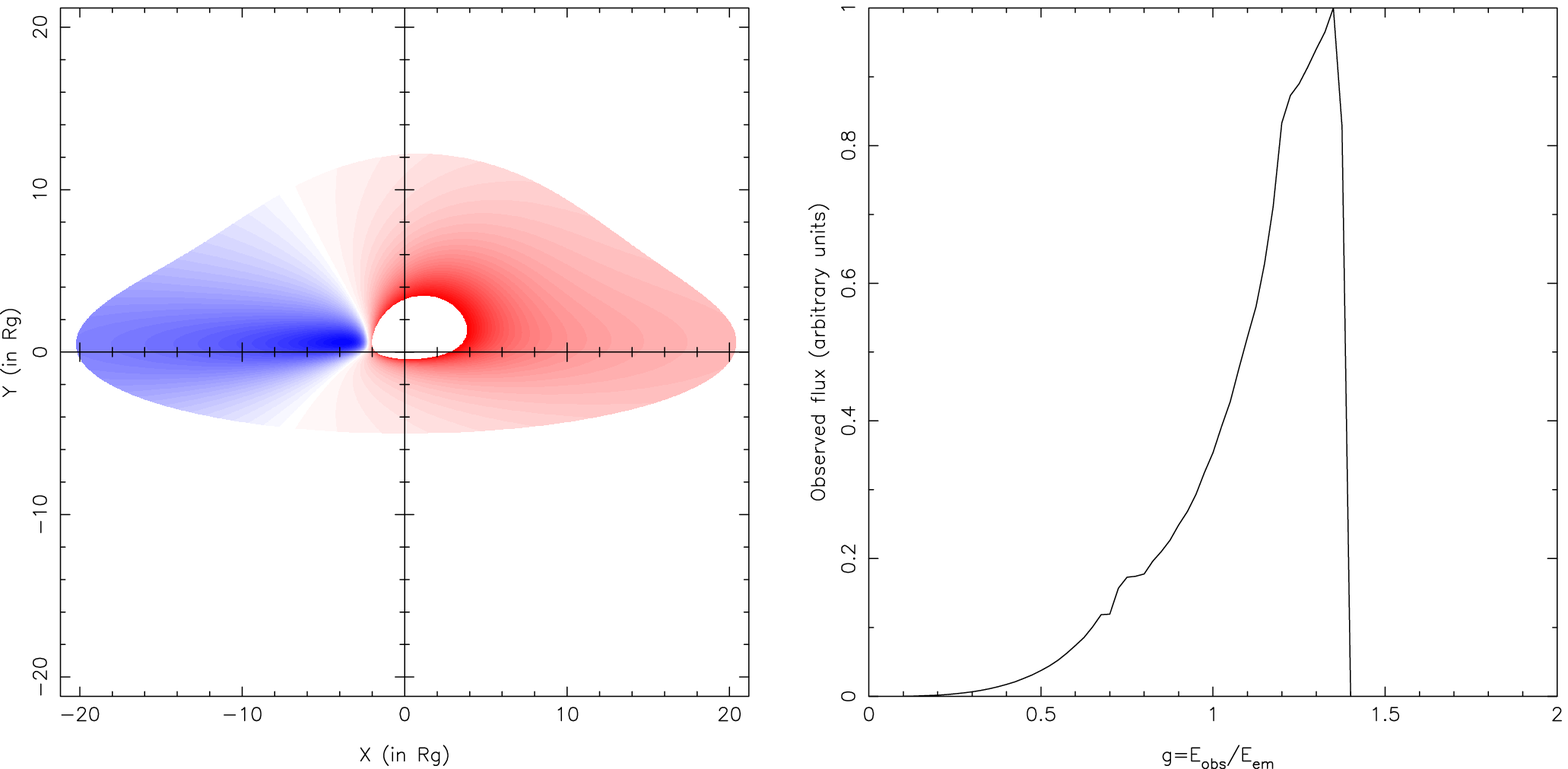}
\caption{The same as in Fig. \ref{fig1} but for a highly inclined disk
with $i=75^\circ$.}
\label{fig2}
\end{figure}

\begin{figure}[ht!]
\centering
\includegraphics[width=0.85\textwidth]{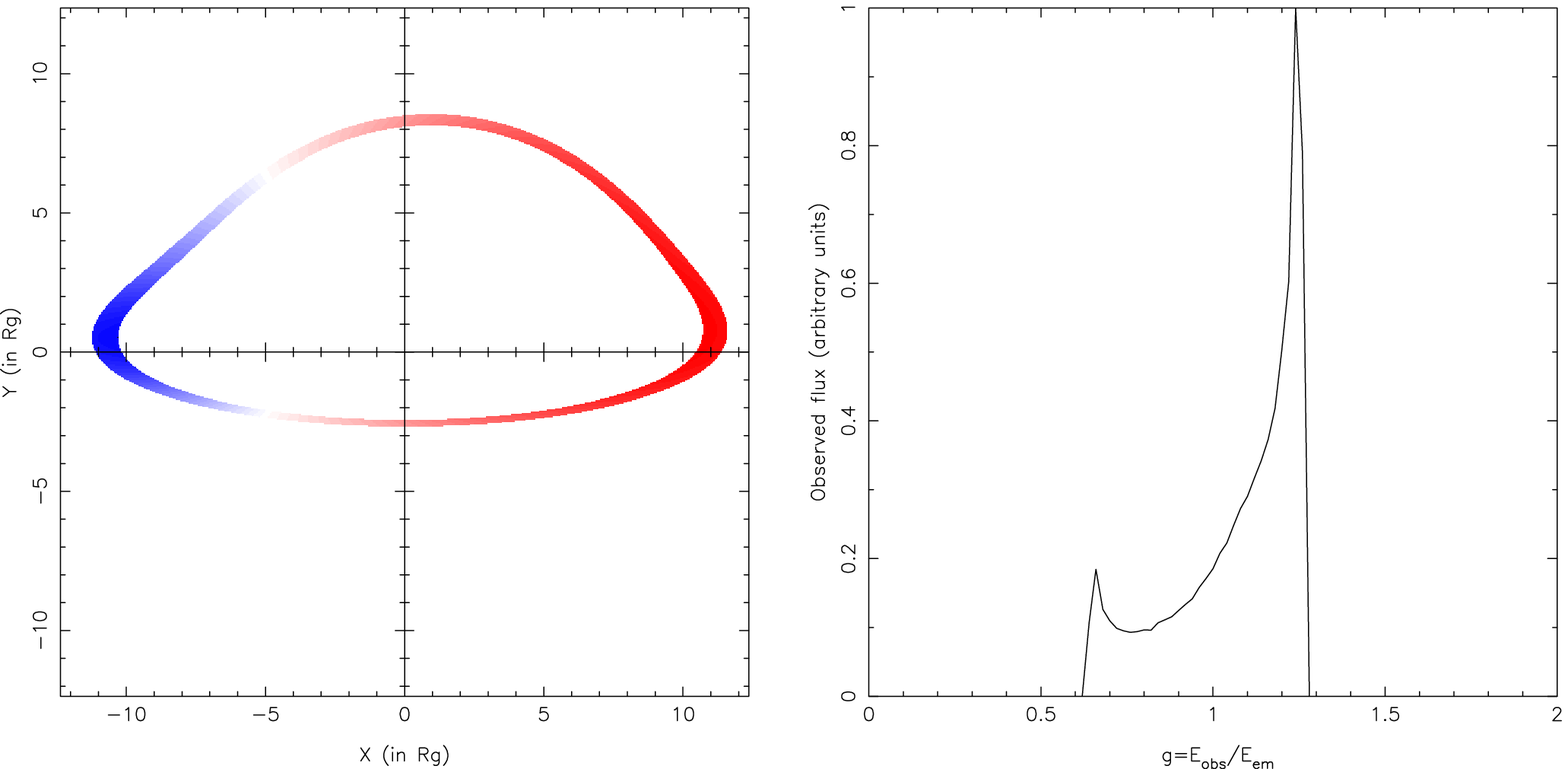} \\
\includegraphics[width=0.85\textwidth]{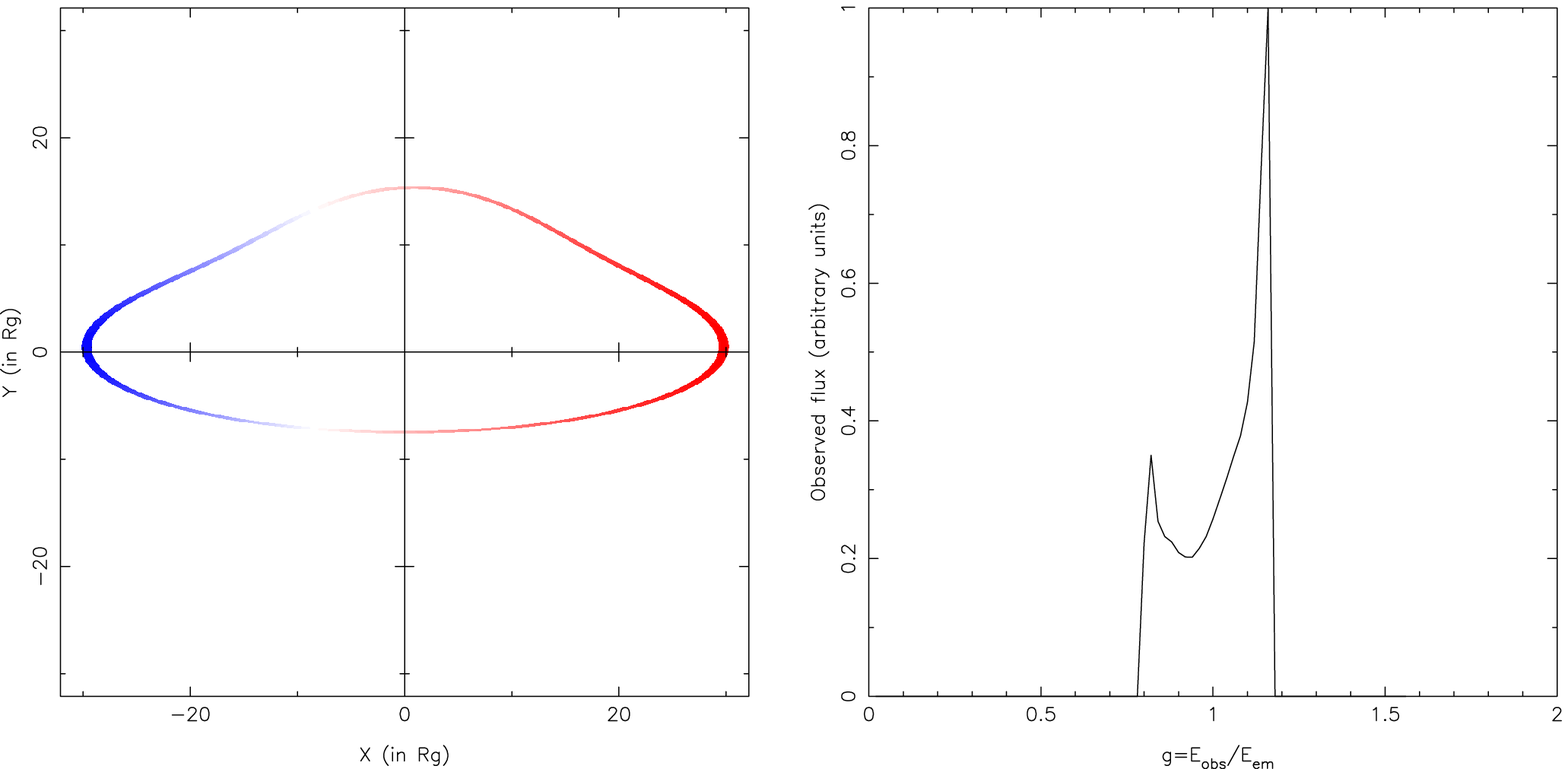} \\
\includegraphics[width=0.85\textwidth]{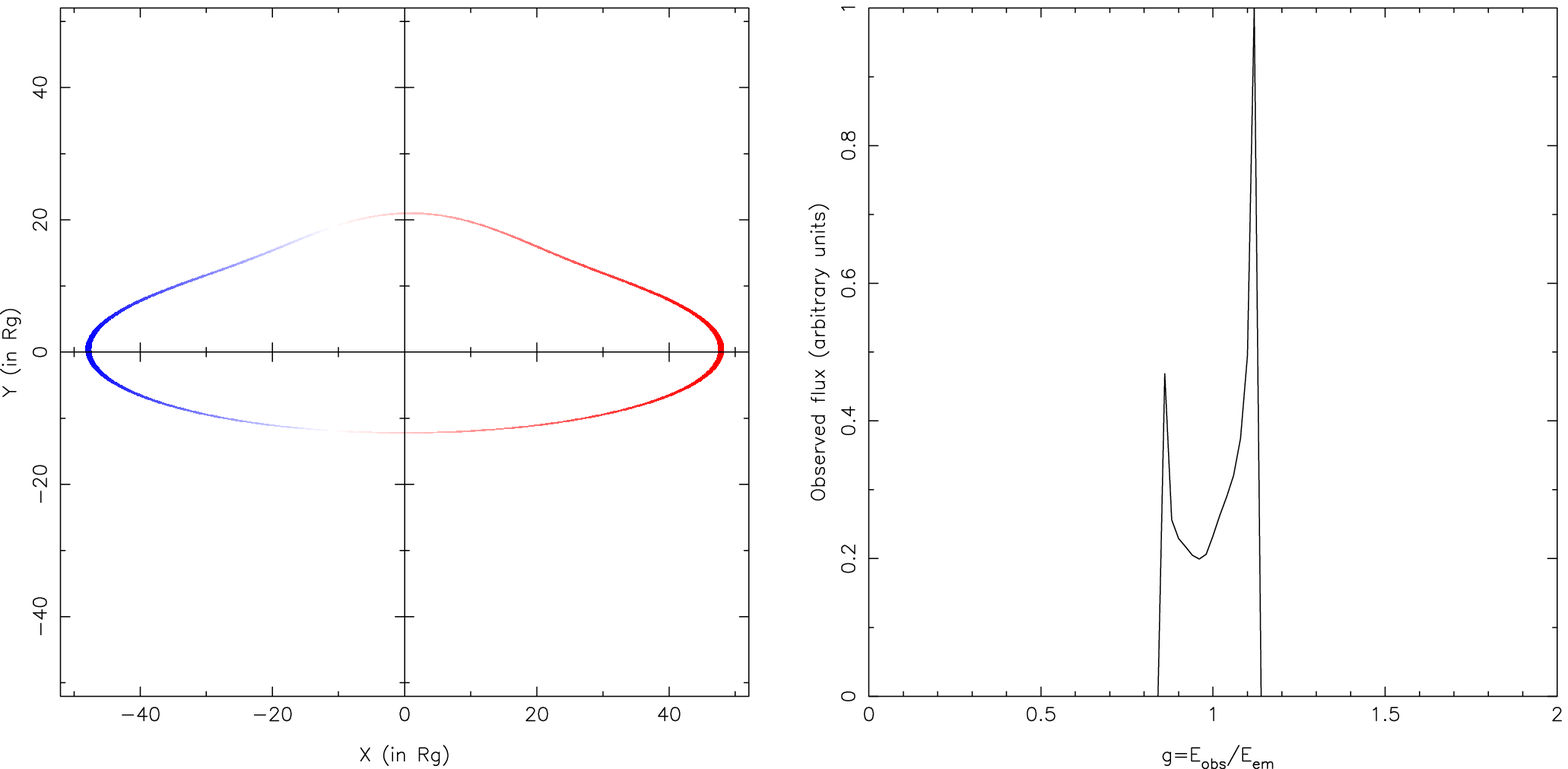}
\caption{The same as in Fig. \ref{fig2} but for the Fe K$\alpha$ line
emitting region in form of narrow annulus with width $=1 R_{g}$, extending from:
$R_{in}=10$ R$_{g}$ to $R_{out}=11$ R$_{g}$ (top),
$R_{in}=30$ R$_{g}$ to $R_{out}=31$ R$_{g}$ (middle) and
$R_{in}=50$ R$_{g}$ to $R_{out}=51$ R$_{g}$ (bottom).}
\label{fig3}
\end{figure}

\section{Numerical simulations}

An accretion disk could have different forms, dimensions and
emissivity, depending on the type of its central BH, whether it is
rotating (Kerr metric) or nonrotating (Schwarzschild metric). We
modeled emission of an accretion disk around supermassive BH using
numerical simulations based on a ray-tracing method in a Kerr
metric, taking into account only photon trajectories reaching the
observer's sky plane in the infinity \cite{pop03a,pop03b,pj06}.
Using this method we are able to obtain colorful images of accretion
disk as would be seen by a distant observer with powerful
high-resolution telescope (see left panels of Figs. \ref{fig1} -
\ref{fig3}). From such disk images we then calculate total observed
flux $F_{obs}$ according to the following expression:
\begin{equation}
F_{obs} \left( {E_{obs}}  \right) = {\int\limits_{image} {\epsilon \left(
{r} \right)}} g^{4}\delta \left( {E_{obs} - gE_{0}}  \right)d\Xi ,
\end{equation}
where $\epsilon \left({r} \right)$ is the disk emissivity, $E_0$ is
the line transition  energy ($E_0^{Fe\ K\alpha}=6.4$ keV),
$g=E_{obs}/E_{em}$ is the energy shift due to relativistic effects
($E_{obs}$ is the observed energy and $E_{em}$ is the emitted energy
from the disk) and $d\Xi$ is solid angle subtended by the accretion
disk on observer's sky. The modeled line profile is then obtained by
binning the flux into the energy shift ($g$) axis (see right panels
of Figs. \ref{fig1} - \ref{fig3}).
 The iron K$\alpha$ line shape strongly depends
on emissivity law of the disk $\epsilon \left({r} \right)$, so we
assume the standard Shakura-Sunyaev disk model \cite{sha73}, where
accretion occurs via an optically thick and geometrically thin disk.

\section{Results}

The effects of strong gravitational field on the Fe K$\alpha$ line
profile have been investigated in order to compare the modeled and
observed line profiles. To obtain modeled line profiles, it is
necessary to define a number of parameters which describe the line
emitting region in the disk, such as constraints for its size, the
disk inclination angle, the mass of the central BH and its angular
momentum. For the disk inclination we adopted the averaged value
from the study of the Fe K$\alpha$ line profiles of 18 Seyfert 1
galaxies: $i=35^\circ$ (see \cite{pop06} and references therein).
The inner radius $R_{in}$ of the disk can not be smaller than the
radius of the marginally stable orbit $R_{ms}$, that corresponds to
$R_{ms}=6R_g$ (gravitational radius $R_g=GM/c^2$, where $G$ is
gravitational constant, $M$ is the mass of central BH, and $c$ is
the velocity of light) in  the Schwarzschild metric and to
$R_{ms}=1.23R_g$ in the case of the Kerr metric with angular
momentum parameter $a=0.998$. To select the outer radius $R_{out}$
of the disk, we take into account some recent investigations of the
Fe K$\alpha$ line profile showing that it should be emitted from the
innermost part of the disk which outer radius is within several tens
of R$_g$ (see \cite{pop06} and references therein).

In order to study observational effects of strong gravity in
vicinity of supermassive BH in the center of AGN, we analyzed three
cases for the Fe K$\alpha$ line emitting region: (i)
$R_{in}=R_{ms}$, $R_{out}=20$ R$_{g}$ and $i=35^\circ$ (in both
Schwarzschild and Kerr metric), (ii) $R_{in}=R_{ms}$, $R_{out}=20$
R$_{g}$ and $i=75^\circ$ (in both Schwarzschild and Kerr metric) and
(iii) $i=75^\circ$ and the line emitting region in Kerr metric with
$a=0.998$ is in form of narrow annulus with width $=1 R_{g}$,
located between: (iiia) $R_{in}=10$ R$_{g}$ and $R_{out}=11$
R$_{g}$, (iiib) $R_{in}=30$ R$_{g}$ and $R_{out}=31$ R$_{g}$ and
(iiic) $R_{in}=50$ R$_{g}$ and $R_{out}=51$ R$_{g}$.

Illustrations of an accretion disk and the corresponding Fe
K$\alpha$ line shapes in the first case for Schwarz\-schild and Kerr
metric are presented in Fig. \ref{fig1}. As one can see in Fig.
\ref{fig1}, the red peak of the Fe K$\alpha$ line is brighter in
case of almost maximally rotating BH, but at the same time it is
also more embedded into the blue peak wing and therefore less
separable from it. Consequently, angular momentum of the central BH
has significant influence on the line shape which supports
assumption that the line originates from the innermost part of
accretion disk, close to the central BH. This fact can be used for
estimation of angular momentum of central BH in observed AGN (see
e.g. \cite{tan95}).

Fig. \ref{fig2} contains illustrations of the line emitting regions
and the corresponding line shapes in the case of highly inclined
disk ($i=75^\circ$). Here, the line profiles are broader than in the
first case, mostly due to higher inclination. As it can be seen in
Fig. \ref{fig2}, in case of Kerr metric, the red peak of the line is
again more embedded into its blue peak wing (as in the first case)
and it confirms that this effect can be most likely attributed to
angular momentum.

Results for the third analyzed case are presented in Fig.
\ref{fig3}. From this figure one can see how the Fe K$\alpha$ line
profile is changing as the function of distance from central BH.
When the line emitters are located at the lower radii of the disk,
i.e. closer to the central BH, the lines are broader and the line
profiles are more asymmetric (see Fig. \ref{fig3}). If the line
emission is originating at larger distances from the BH, the red
peak of the line becomes brighter and line profile narrower and more
symmetric. In majority of AGN, where the broad Fe K$\alpha$ line is
observed\footnote{Note here that in some AGN only the narrow Fe
K$\alpha$ line is observed, but it is supposed to be emitted in the
disk corona that is located farther from the disk, and therefore,
these relativistic effects cannot be detected in the line profile},
its profile is more similar to the modeled profile as obtained under
assumption that the line emitters are located close to the central
BH. Therefore, comparisons between the observed and modeled Fe
K$\alpha$ line profiles can bring us some essential information
about strong gravitational field in vicinity of central supermassive
BH of AGN.

\section{Conclusions}

We performed numerical simulations based on a ray-tracing method in
a Kerr metric in order to model the emission of accretion disk
around supermassive BH of AGN. We also simulated the influence of a
strong gravitational field on the Fe K$\alpha$ line, showing that
these effects can be detected in the observed line shapes. According
to the obtained results, angular momentum or spin of central
supermassive BH of AGN has significant influence on the line
profile. Therefore, the analysis of the high resolution observations
of the Fe K$\alpha$ line could be used for determination of the
space-time geometry (metric) in vicinity of the supermassive BH,
supposed to be in heart of AGN.

\begin{acknowledgement}
This work is a part of the project (146002) "Astrophysical
Spectroscopy of Extragalactic Objects" supported by the Ministry of
Science of Serbia. L. \v C. Popovi\' c is supported by Alexander von
Humboldt foundation through Fritz Thyssen Special Programme.
\end{acknowledgement}

\end{document}